\documentclass[aps,pra,twocolumn,superscriptaddress]{revtex4-1}

\usepackage{graphicx,subfigure}
\usepackage{amsmath,amssymb,bm,amsthm}
\usepackage{mathtools}
\usepackage{multirow}
\usepackage{textcomp}
\usepackage{float}
\usepackage{xcolor}
\usepackage[normalem]{ulem}
\usepackage{dsfont}
\usepackage{url}

\newcommand{\ave}[1]{\left \langle #1 \right \rangle}
\newcommand{\ket}[1]{\left | #1 \right \rangle}
\newcommand{\bra}[1]{\left \langle #1 \right |}
\newcommand{\mele}[3]{\left \langle #1 \middle | #2 \middle | #3 \right \rangle}
\newcommand{\inp}[2]{\left \langle #1 \middle | #2 \right \rangle}
\newcommand{\tr}[1]{\text{Tr} \left [ #1 \right ]}

\begin{document}

\title{Automated discovery of autonomous quantum error correction schemes}

\author{Zhaoyou Wang}
\affiliation{E. L. Ginzton Laboratory and the Department of Applied Physics, Stanford University, Stanford, CA 94305 USA}
\author{Taha Rajabzadeh}
\affiliation{Department of Electrical Engineering, Stanford University, Stanford, CA 94305 USA}
\author{Nathan Lee}
\author{Amir H. Safavi-Naeini}
\affiliation{E. L. Ginzton Laboratory and the Department of Applied Physics, Stanford University, Stanford, CA 94305 USA}

\date{\today}

\begin{abstract}
We can encode a qubit in the energy levels of a quantum system. Relaxation and other dissipation processes lead to decay of the fidelity of this stored information. Is it possible to preserve the quantum information for a longer time by introducing additional drives and dissipation? 
The existence of autonomous quantum error correcting codes answers this question in the positive. Nonetheless, discovering these codes for a real physical system, i.e., finding the encoding and the associated driving fields and bath couplings, remains a challenge that has required intuition and inspiration to overcome. In this work, we develop and demonstrate a computational approach based on adjoint optimization for discovering autonomous quantum error correcting codes given a description of a physical system. 
We implement an optimizer that searches for a logical subspace and control parameters to better preserve quantum information. We demonstrate our method on a system of a harmonic oscillator coupled to a lossy qubit, and find that varying the Hamiltonian distance in Fock space -- a proxy for the control hardware complexity -- leads to discovery of different and new error correcting schemes. We discover what we call the $\sqrt{3}$ code, realizable with a Hamiltonian distance $d=2$, and propose a hardware-efficient implementation based on superconducting circuits.
\end{abstract}

\maketitle

Physical qubits always live in noisy environments which leads to decoherence and hinders the development of scalable quantum computers.
Quantum error correction (QEC) solves this problem by encoding logical states in a way that errors caused by the environment can be detected and corrected without accessing the encoded quantum information~\cite{Nielsen2010a}.
The standard implementation of QEC involves error syndrome measurements followed by adaptive recovery operations~\cite{Nielsen2010a}, with the drawback of introducing additional errors caused by imperfect measurements~\cite{Ofek2016,Hu2019,Campagne-Ibarcq2020} and significant hardware overhead associated with the real-time classical feedback~\cite{Fowler2012,Terhal2015}. In contrast, autonomous quantum error correction (AQEC) circumvents the necessity of classical adaptive control by embedding the active measurement and feedback processes into the passive internal dynamics of the system~\cite{Kapit2016a,Gertler2021,Lebreuilly2021,Kerckhoff2010a}. With an engineered interaction between the qubit and a lossy ancilla, accumulated entropy in the qubit due to physical errors can be coherently transferred to the ancilla in real-time and then evacuated through ancilla decay~\cite{Kapit2016a,Gertler2021,Lebreuilly2021}.

An experimental platform will feature limited hardware control and more complex errors, which sometimes deviate from the assumptions underlying many QEC codes. Designing the optimal platform-specific QEC scheme is therefore highly nontrivial and will likely require a numerical approach that takes into account the hardware constraints.
In this direction~\cite{Leung1997,Reimpell2005,Fletcher2008,Kosut2009}, it has been  shown that the encoding and decoding operations can be adapted to a given error channel via iterative convex optimization~\cite{Reimpell2005,Kosut2009}, assuming arbitrary operations are achievable.
More recently, quantum gate based methods were developed to incorporate certain features at the physical level, including the available gate set and qubit connectivity~\cite{Fosel2018,Johnson2017}.
It is important to extend these approaches to take into account  decoherence during the gate operation and measurements, as well as any coherent leakage out of the qubit computational subspace~\cite{Chen2016a}.
AQEC on the other hand, only requires Hamiltonian dynamics and ancilla relaxation to preserve the quantum information. Therefore automated discovery of AQEC schemes naturally incorporates the native device Hamiltonian and decay channels, bringing it closer to experimental deployment.

In this paper, we develop a numerical framework (\texttt{AutoQEC}) for automatically designing AQEC schemes for a given experimental platform.
\texttt{AutoQEC} aims to discover strategies that preserve the encoded quantum information by optimizing over the logical states and control parameters.
We demonstrate \texttt{AutoQEC} on a system consisting of a single bosonic mode coupled to a lossy qubit, and find that the resulting AQEC scheme depends on the constraints on the system, such as dissipation channels and Hamiltonian connectivity. Finally, we propose a circuit implementation of the discovered AQEC scheme. 

\begin{figure}[h]
    \centering
    \includegraphics[width=0.48\textwidth]{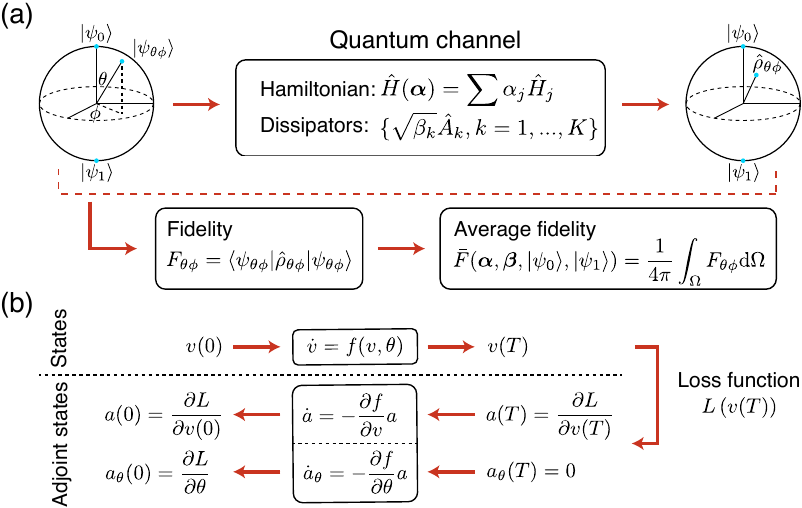}
    \caption{Schematic of the average fidelity and the adjoint method. (a) An arbitrary state $\ket{\psi_{\theta\phi}}$ on the Bloch sphere is mapped to a density matrix $\hat{\rho}_{\theta\phi}$ by a quantum channel described by a set of control Hamiltonians and dissipators. Averaging the single state fidelity $F_{\theta\phi}$ over the Bloch sphere gives the average fidelity $\bar{F}$. (b) A loss function $L(v(T))$ depending on the final state $v(T)$ of an ODE will implicitly be a function of the initial state $v(0)$ and the ODE parameter $\theta$. The gradients of $L$ with respect to $v(0)$ and $\theta$ can be computed by integrating the adjoint equations backward in time.}
    \label{fig1}
\end{figure}

Our method is illustrated in Fig.~\ref{fig1}(a) where the physical system is described by a parametrized Hamiltonian $\hat{H} (\bm{\alpha}) = \sum_{j=1}^N \alpha_j \hat{H}_j$ and a set of dissipators $\{\sqrt{\beta_k} \hat{A}_k, k=1,...,K\}$. Here $\bm{\alpha}=\{\alpha_j\},\bm{\beta}=\{\beta_k\}$ are the control parameters while $\{\hat{H}_j\}$ and $\{\hat{A}_k\}$ represent the available control Hamiltonians and decay channels in the system.
The quantum dynamics follows the master equation
\begin{equation}\label{eq:master}
    \dot{\hat{\rho}}=-i[\hat{H}(\bm{\alpha}), \hat{\rho}] + \sum_{k=1}^{K} \beta_k \left( \hat{A}_k \hat{\rho} \hat{A}_{k}^{\dagger} - \frac{1}{2} \left \{ \hat{A}_{k}^{\dagger}\hat{A}_{k}, \hat{\rho} \right\}\right).
\end{equation}
Learning a logical qubit is equivalent to finding a 2-dimensional subspace that is well-protected under the dynamics of Eq.~(\ref{eq:master}).
More concretely, starting from an arbitrary point on the Bloch sphere spanned by basis vectors $\ket{\psi_0}$ and $\ket{\psi_1}$ as the initial state (Fig.~\ref{fig1}(a))
\begin{equation}
    \ket{\psi_{\theta \phi}} = \cos \frac{\theta}{2} \ket{\psi_0} + e^{i\phi} \sin \frac{\theta}{2} \ket{\psi_1} ,
\end{equation}
the fidelity $F_{\theta\phi} = \mele{\psi_{\theta \phi}}{\hat{\rho}_{\theta \phi}}{\psi_{\theta \phi}}$ characterizes how much information is preserved for $\ket{\psi_{\theta \phi}}$, where $\hat{\rho}_{\theta \phi}$ is the state at some later time $T$ evolved under Eq.~(\ref{eq:master}).
By integrating the single state fidelity over the Bloch sphere $\Omega$, the average fidelity of the logical subspace is defined as
\begin{equation}\label{eq:avg_F}
    \bar{F}(\bm{\alpha},\bm{\beta},\ket{\psi_0},\ket{\psi_1}) \equiv \frac{1}{4\pi} \int_{\Omega} F_{\theta\phi} \text{d} \Omega.
\end{equation}
We found that using $\bar{F}$ as the objective function in \texttt{AutoQEC} is sometimes too constraining and leads to an untenable optimization landscape with high sensitivity to the parameters. This can be understood by considering how a small change in the energy levels can cause a phase to build up over the evolution time and make the fidelity fluctuate rapidly with the optimization parameters. Our solution was to find the best overlap modulo a $Z$ rotation in the logical subspace (see SI). A similar but even less constraining approach has been used previously where the recoverable quantum information is maximized~\cite{Fosel2018}.  

Optimization of the average fidelity $\bar{F}$ is in general a high-dimensional non-convex problem, and local gradient information could potentially accelerate the search of good solutions.
The gradients of $\bar{F}$ with respect to the logical states $\{ \ket{\psi_0},\ket{\psi_1} \}$ and the control parameters $\{ \bm{\alpha},\bm{\beta} \}$ can be calculated with the adjoint method~\cite{Pontryagin2018,Chen2018a}, a technique for efficiently backpropagating gradients through an ordinary differential equation (ODE).
Mathematically, considering a state $v$ obeying the ODE $\dot{v} = f(v,\theta)$ with parameter $\theta$ and a loss function $L(v(T))$ depending on the final state at time $T$, the key quantity introduced in the adjoint method is the adjoint state $a(t) = \partial L / \partial v(t)$ whose dynamics satisfies $\dot{a} = -\frac{\partial f}{\partial v} a$ (Fig.~\ref{fig1}(b)).
Therefore the gradient of $L$ with respect to the initial state $a(0)=\partial L / \partial v(0)$ can be computed by starting from $a(T) = \partial L / \partial v(T)$ and integrating the adjoint equation backward in time.
Similarly, the gradient with respect to the ODE parameter $\theta$ can be computed by solving another adjoint equation $\dot{a}_\theta = -\frac{\partial f}{\partial \theta} a$ backward in time starting from $a_\theta (T) = 0$ (Fig.~\ref{fig1}(b)). 
Notice that even though we use scalar notations here for simplicity, the results above can be easily generalized to the vector case (see SI).

\begin{figure}[t]
    \centering
    \includegraphics[width=0.48\textwidth]{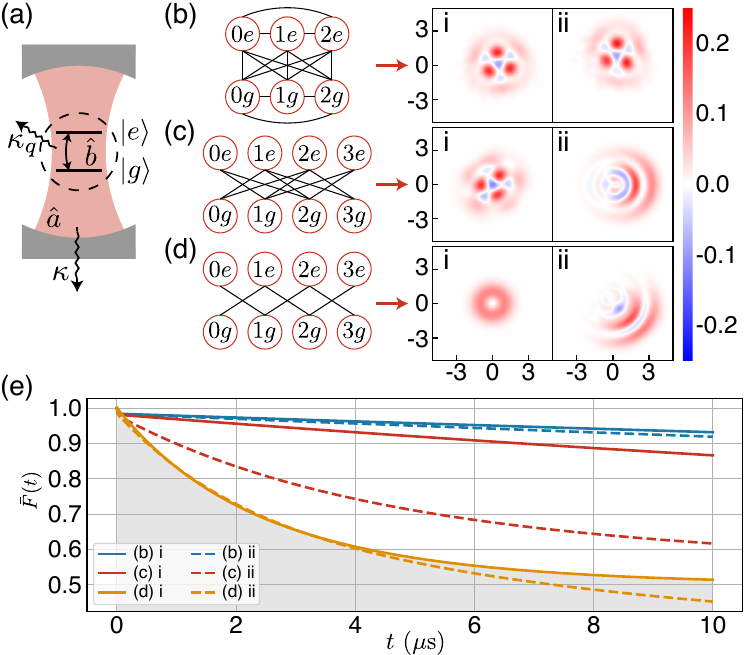}
    \caption{Optimization results for different Hamiltonian distances. (a) Schematic of a single bosonic mode $\hat{a}$ coupled to a lossy ancilla qubit $\hat{b}$. (b-d) Wigner functions of the discovered codes for all-to-all (b), distance $d=2$ (c) and $d=1$ (d) Hamiltonians. (e) Average fidelity for all-to-all (blue), $d=2$ (red) and $d=1$ (yellow) results. The grey shade region indicates fidelities below or equal to the break-even point.}
    \label{fig2}
\end{figure}

We use \texttt{AutoQEC} to study autonomous implementations of bosonic codes~\cite{Gottesman2001,Leghtas2013b,Michael2016,Albert2018,Ofek2016,Hu2019,Gertler2021}, motivated by their advantages of hardware efficiency and simplified error models over the traditional qubit based QEC~\cite{Steane1996a,Nielsen2010a,Fowler2012}.
The system consists of a single harmonic oscillator as the storage mode of quantum information coupled to a lossy ancilla qubit for entropy evacuation (Fig.~\ref{fig2}(a)), and the dissipators are $\sqrt{\kappa} \hat{a}$ and $\sqrt{\kappa_q} \hat{b}$ where $\hat{a}$ and $\hat{b}$ are the annihilation operators for the bosonic mode and the qubit respectively.
We choose the basis states of the joint system as $\ket{n,g(e)}$ where $n$ is the index of the Fock state and $g(e)$ represents the qubit in its ground (excited) state.
During the optimization of $\bar{F}$, only the logical states $\{ \ket{\psi_0},\ket{\psi_1} \}$ and the Hamiltonian parameters $\bm{\alpha}$ are updated while the loss rates are fixed at $\kappa/2\pi = 0.1$~MHz and $\kappa_q/2\pi = 20$~MHz.
We choose a total evolution time $T = 0.5~\mu$s, and bound the driving strength of each Hamiltonian by $|\alpha_j|/2\pi \leq 10~$MHz reflecting what we consider as realistic couplings rates. 

We begin by considering Hamiltonians with all-to-all coupling in the Hilbert space (Fig.~\ref{fig2}(b)) where the control Hamiltonians $\{\hat{H}_j\}$ include couplings between any two basis states $\{\ket{m,g(e)} \bra{n,g(e)}\}$. Running the \texttt{AutoQEC} optimizer on this problem leads repeatabily to codes such as those shown in Fig.~\ref{fig2}(b)i and ii. In both cases, we plot the Wigner function of the maximally mixed state $\hat{\rho}_{\text{code}} = \frac{1}{2} (\ket{\psi_0} \bra{\psi_0} + \ket{\psi_1} \bra{\psi_1} )$
as a basis-independent representation of the logical subspace. The two logical subspaces are basically equivalent to each other up to a random displacement and rotation in phase space.
Moreover, the code in Fig.~\ref{fig2}(b)i is almost identical ($F \equiv 2\tr{\hat{\rho}_{1} \hat{\rho}_{2}} \approx99.4\%$, where $\hat{\rho}_{1}$ and $\hat{\rho}_{2}$ are the two codes) with the $\sqrt{17}$ code~\cite{Michael2016,Albert2018}, which is the smallest code in photon number that allows exact correction of a single photon loss error.
The time-evolution of the average fidelity quantifies the AQEC performance. We plot this for both resulting codes (Fig.~\ref{fig2}(e) blue solid and dashed lines), and see that it exceeds the average fidelity for the trivial encoding (Fig.~\ref{fig2}(e) grey shade region boundary), \emph{i.e.}, the $\ket{0}$ and $\ket{1}$ subspace with $\hat{H}=0$ which defines break-even.

To avoid the significant experimental challenge of implementing an all-to-all Hamiltonian for a harmonic oscillator, we consider a more realizable restricted set of control Hamiltonians containing only terms $\{\ket{m,g}\bra{n,e}\}$ (and the conjugates) where $0<|m-n|\leq d$ and the distance $d=2$ (Fig.~\ref{fig2}(c)). We choose such a restriction since we expect interaction terms such as $\hat{a}^\dagger \hat{b}$ and $\hat{a}^{\dagger 2} \hat{b}$ to be selectively realizable by engineering the drive frequencies given large dispersive couplings (see implementation details below).
We also avoid coupling terms like $\{\ket{m,g}\bra{n,g}\}$ and $\{\ket{m,e}\bra{n,e}\}$ since \emph{independently} engineering them for a linear system $\hat{a}$ is difficult as we no longer have access to the dispersive nonlinearity.
Over many runs of \texttt{AutoQEC}, we discover two different types of results that exceed break-even.
Fig.~\ref{fig2}(c)i shows the logical subspace of a discovered bosonic code, which exhibits error correction performance (Fig.~\ref{fig2}(e) red solid line) that approaches the all-to-all coupling results.
A second encoding shown in Fig.~\ref{fig2}(c)ii only provides partial protection in the logical subspace, where $\ket{\psi_0}$ only occupies low photon number states and $\ket{\psi_1}$ only occupies high photon number states (see SI). In a way that is reminiscent of recent works on error-biased cat qubits~\cite{Guillaud2019}, both $\ket{\psi_0}$ and $\ket{\psi_1}$ are preserved by the Hamiltonian with fidelities above break-even, but not some of their superpositions (see SI). Nevertheless, the average fidelity of this encoding still exceeds break-even (Fig.~\ref{fig2}(e) red dashed line).

We fail to discover any error correcting codes with performance beyond break-even with a Hamiltonian distance $d=1$ (Fig.~\ref{fig2}(d)). Most \texttt{AutoQEC} searches end with the $\ket{0}$ and $\ket{1}$ subspace (Fig.~\ref{fig2}(d)i), while occasionally we also obtain states (Fig.~\ref{fig2}(d)ii) similar to the $d=2$ case Fig.~\ref{fig2}(c)ii, with a fidelity (Fig.~\ref{fig2}(e) yellow dashed line) slightly below break-even.

\begin{figure}[t]
    \centering
    \includegraphics[width=0.48\textwidth]{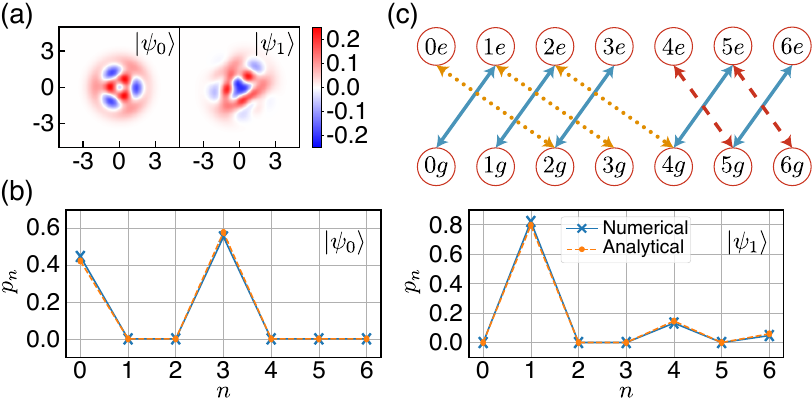}
    \caption{The $\sqrt{3}$ code. (a) Wigner functions of the two logical basis states from \texttt{AutoQEC}. (b) The photon number distribution of the analytically derived $\sqrt{3}$ code agrees well with the numerical result, where $\ket{\psi_0} \in \text{span}\{ \ket{0}, \ket{3} \}$ and $\ket{\psi_1} \in \text{span}\{ \ket{1}, \ket{4}, \ket{6} \}$. (c) Required Hamiltonian couplings for autonomously protecting the code.}
    \label{fig3}
\end{figure}

We further investigate the $d=2$ code in Fig.~\ref{fig2}(c)i. Figure~\ref{fig3} shows two orthogonal code words spanning the logical subspace and the associated Hamiltonian.
Inspired by the numerical results (Fig.~\ref{fig3}(a-b)), we analytically derive the logical states ($F\approx99.9\%$; see SI) and name it the $\sqrt{3}$ code, after the average number of photons in its codewords.
The Hamiltonian found by \texttt{AutoQEC} uses three types of couplings (Fig.~\ref{fig3}(c)): type 1 (blue solid line) $\ket{n-1,g} \leftrightarrow \ket{n,e}$, type 2 (red dashed line) $\ket{n+1,g} \leftrightarrow \ket{n,e}$, and type 3 (yellow dotted line) $\ket{n+2,g} \leftrightarrow \ket{n,e}$. The general form of this Hamiltonian is 
\begin{equation}\label{eq:AQEC_H}
    \hat{H} = \sum_{l=1,2,3}\hat{H}^{(l)} , \quad \hat{H}^{(l)} = \sum_n \alpha^{(l)}_n \ket{n} \bra{n+d_l} \otimes \hat{b}^\dagger + \text{h.c.}
\end{equation}
where $\hat{H}^{(l)}$ corresponds to the type $l$ couplings with distances $d_1=-1,d_2=1,d_3=2$ and $\hat{b} = \ket{g} \bra{e}$.

Our goal now is to find a physical implementation of the Hamiltonian Eq.~(\ref{eq:AQEC_H}) discovered by \texttt{AutoQEC}. In circuit QED, a method for selectively driving such transitions uses off-resonant coupling between a resonator and a nonlinear ancilla qubit circuit to realize a dispersive $\chi \hat{a}^\dagger \hat{a} \hat{b}^\dagger \hat{b}$ interaction.
In presence of this nonlinear level structure, coupling $\ket{m,g}$ and $\ket{n,e}$ together is achieved by driving the system so that the operator $\hat{a}^{m-n} \hat{b}^\dagger$ (if $m>n$, otherwise $\hat{a}^{\dagger (n-m)} \hat{b}^\dagger$) appears in the Hamiltonian, oscillating at frequency $n\chi$ in the rotating frame of both $\hat{a}$ and $\hat{b}$. 
This approach requires that the driving strength $|\alpha_j|$ is sufficiently weaker than $\chi$ so that a rotating wave approximation (RWA) may be made to drop the undesired couplings.
For example,  $\hat{H}^{(1)}$ is effectively implemented by $\hat{H}_d^{(1)}(t) = f_1(t) \hat{a}^\dagger \hat{b}^\dagger + \text{h.c.}$ with $f_1(t) = \sum_n \alpha^{(1)}_n e^{-i n\chi t} /\sqrt{n}$.
Similarly $\hat{H}^{(2)}$ and $\hat{H}^{(3)}$ can be realized by including driving fields that properly modulate the $\hat{a}\hat{b}^\dagger$ and $\hat{a}^2\hat{b}^\dagger$ terms of the Hamiltonian (see SI).

\begin{figure}[t]
    \centering
    \includegraphics[width=0.48\textwidth]{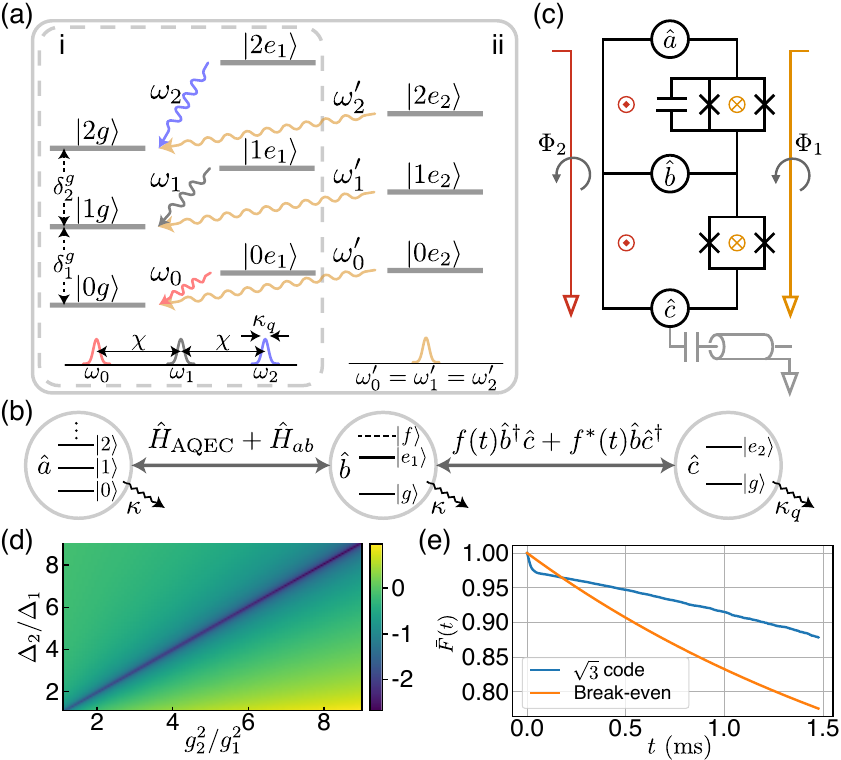}
    \caption{Circuit implementation of the discovered AQEC scheme. (a) Level diagram and emission spectrum before (i) and after (ii) erasing the which-way information. (b) Schematic interactions within the extended system with an additional lossy qubit $\hat{c}$. (c) Circuit design that implements the desired interactions by modulating the flux threading the smaller (yellow) and larger (red) loops. (d) $\log_{10} \mathcal{B}$ as a function of $g_2^2/g_1^2$ and $\Delta_2/\Delta_1$, which is minimized when $g_1^2/\Delta_1 \approx g_2^2/\Delta_2$. The simulation is performed with $\Delta_1/2\pi=1~$GHz and $g_1/2\pi=100~$MHz. (e) Average fidelity of the $\sqrt{3}$ code from a time-domain simulation of the full circuit dynamics.}
    \label{fig4}
\end{figure}

Before even considering the circuit implementation, we note that the above approach has a serious shortcoming related to how the dispersive nonlinearity modifies the dissipative dynamics.
The photon emitted by the relaxation $\ket{n,e} \rightarrow \ket{n,g}$ at frequency $\omega_n$ is spaced from $\omega_{n \pm 1}$ by $\chi$ (Fig.~\ref{fig4}(a)i), which is much larger than the linewidth $\kappa_q \sim |\alpha_j|$ of these emission lines due to the RWA requirement $\chi \gg |\alpha_j|$.
Therefore the emitted photons would leak information about the photon number distribution and
the correct dissipator is no longer $D[\hat b]$ for which the code was optimized, but an incoherent sum of terms $D[|n,g\rangle\langle n,e|]$.  

We erase the which-way information by including an additional lossy ancillary system and a number of drives with frequencies tuned to overlap the spectrum of emitted photons. This extra lossy qubit $\hat{c}$ does not dispersively couple to $\hat{a}$ -- in contrast to qubit $\hat{b}$, which we also now make long-lived as it mediates interaction between the storage mode $\hat{a}$ and the lossy ancilla $\hat{c}$ (Fig.~\ref{fig4}(b)).
Let $\ket{e_1}$ and $\ket{e_2}$ be the excited states of qubits $\hat{b}$ and $\hat{c}$ respectively, the interaction $\hat{H}_{bc}$ between $\hat{b}$ and $\hat{c}$ should quickly transfer any occupation in $\ket{n, e_1}$ caused by a photon loss error in $\hat{a}$ into $\ket{n, e_2}$, which then relaxes back to $\ket{n, g}$.
Due to the absence of dispersive coupling, the emitted photon from $\hat{c}$ doesn't reveal information about the $\hat{a}$ states which eliminates the which-way information (Fig.~\ref{fig4}(a)ii).
Because the transition frequency between $\ket{n, e_1}$ and $\ket{n, e_2}$ depends on $n$, the desired swapping Hamiltonian $\hat{H}_{bc}=\Omega \sum_n e^{-in\chi t} \ket{n,e_1}\bra{n,e_2} + \text{h.c.}$ is time-dependent, which can be implemented as $f(t) \hat{b}^\dagger \hat{c} + \text{h.c.}$ with $f(t) = \Omega\sum_n e^{-in\chi t}$ and $\chi \gg \Omega$.

Having incorporated the second ancillary qubit, we propose a superconducting circuit (Fig.~\ref{fig4}(c)) that implements the discovered AQEC scheme.
The direct capacitive coupling between $\hat{a}$ and $\hat{b}$ gives rise to a linear interaction $g(\hat{a}^\dagger\hat{b} + \hat{a} \hat{b}^\dagger)$ which effectively generates a dispersive coupling $\chi \hat{a}^\dagger \hat{a} \hat{b}^\dagger \hat{b}$ with $\chi = 4 g^2 /\Delta$ in the large detuning regime $\Delta = |\omega_a - \omega_b| \gg g$~\cite{James2007,Reiter2012a}.
The two SQUIDs introduce nonlinear interactions of the form $\cos \left( \varphi_x (\hat{x}+\hat{x}^\dagger) + \varphi_y (\hat{y}+\hat{y}^\dagger) \right)$ and $\sin \left( \varphi_x (\hat{x}+\hat{x}^\dagger) + \varphi_y (\hat{y}+\hat{y}^\dagger) \right)$ with $(x,y)=(a,b),(b,c)$.
Different terms from the $\cos$ and $\sin$ Taylor expansions can be selected by parametrically driving the flux threading the circuit loops~\cite{Kapit2016a}, which realize both the AQEC Hamiltonian between $\hat{a}$ and $\hat{b}$ and the swapping Hamiltonian $\hat{H}_{bc}$.
With proper placement of mode frequencies, the $\hat{a} \leftrightarrow \hat{b}$ and $\hat{b} \leftrightarrow \hat{c}$ couplings can be controlled at very different driving frequencies and only a minimum of two flux lines are required (see SI).

The interaction $g(\hat{a}^\dagger\hat{b} + \hat{a} \hat{b}^\dagger)$ of the proposed circuit implements the needed dispersive coupling at the leading order. However, its higher order effects induce a Kerr nonlinearity on the $\hat{a}$ mode~\cite{Zhang2021} that can be comparable to $\kappa$, causing real deleterious effects on the coherence of the system.
Due to this Kerr nonlinearity, when the $\hat{a}$ mode decays from $\ket{n,g}$ to $\ket{n-1,g}$, the emitted photon frequencies $\{\delta_n^{g}\}$ are not perfectly identical, but rather have a finite bandwidth $\mathcal{B} \equiv \max \{\delta_n^{g}\} - \min \{\delta_n^{g}\}$. For realistic parameters where $\mathcal{B}>\kappa$, the environment  learns about the photon number distribution of the resonator, inducing  additional dephasing. \texttt{AutoQEC}'s assumption of a dissipator $D[\hat a]$ is then no longer correct and an incoherent sum of $D[|n-1,g\rangle\langle n,g|]$ more accurately captures the resonator damping process, reducing the fidelity.

Properly addressing parasitic Kerr nonlinearities is a common challenge in bosonic quantum hardware~\cite{Zhang2021}. By leveraging the $\ket{f}$ level of the qubit $\hat{b}$ and engineering the linear coupling between $\hat{a}$ and $\hat{b}$, we find a way to suppress the emission bandwidth $\mathcal{B}$ by a few orders of magnitude.
Consider the Hamiltonian of a harmonic oscillator coupled to a three-level qubit
\begin{equation}\label{eq:H_ab}
    \begin{split}
        \hat{H}_{ab} =& \Delta_1 \ket{e} \bra{e} + \Delta_2 \ket{f} \bra{f} + g_1 (\hat{a}^\dagger \ket{g}\bra{e} + \hat{a}\ket{e}\bra{g}) \\
        & + g_2 (\hat{a}\ket{f}\bra{e} + \hat{a}^\dagger \ket{e}\bra{f}).
    \end{split}
\end{equation}
We find that by judiciously setting the detunings $\Delta_1$ and $\Delta_2$ such that $g_1^2/\Delta_1 \approx g_2^2/\Delta_2$,  $\mathcal{B}$ is eliminated to the leading order (see SI).
This is also verified numerically in Fig.~\ref{fig4}(d) where $\log_{10} \mathcal{B}$ as a function of $g_2^2/g_1^2$ and $\Delta_2/\Delta_1$ is minimized along the diagonal.

Finally, we perform a full quantum simulation of the proposed circuit. Given our choice of $g$ and the detunings, an experimentally realistic $\chi/2\pi \approx 10~$MHz is achieved. To safely satisfy the parameter hierarchy $\chi \gg |\alpha_i| \gg \kappa$, we rescale all control parameters down to $|\alpha_i|/2\pi \approx 10$~kHz and $\kappa/2\pi=0.1$~kHz which has the same AQEC performance but at a different time scale.
A time-domain simulation of the full circuit dynamics (see SI) proves that the average fidelity for the $\sqrt{3}$ code indeed exceeds the break-even point (Fig.~\ref{fig4}(e)) which confirms our circuit design and the AQEC behavior.
Even though $\kappa/2\pi=0.1$~kHz is quite demanding experimentally, we expect it to be improved with better circuit designs and proper compensation of level Stark shifts, and eventually realizable with 3D microwave cavity~\cite{Reagor2016a} or quantum acoustic platforms~\cite{Arrangoiz-Arriola2019a,MacCabe2020}.
We have proposed and demonstrated a numerical framework for automated discovery of AQEC schemes. With future progress on speeding up the physical simulation, extension and application of our method in larger systems may enable learning multiple logical qubits as well as quantum gates in the few logical qubit regime. It would also be important to incorporate more of the physical layer design, which were done ``by hand'' in this work, into the optimization algorithm. Finally, applying the adjoint method to other physical systems with the goal of improving quantum simulation, sensing and communications protocols seems within reach.

\begin{acknowledgments}
This work was funded by the U.S. government through the Office of Naval Research (ONR) under grant No. N00014-20-1-2422, a MURI grant from the U. S. Air Force Office of Scientific Research (Grant No. FA9550-17-1-0002), and the Department of Energy through Grant No. DE-SC0019174. The authors wish to thank NTT Research for their financial and technical support. A.H.S.-N. acknowledges support from the David and Lucille Packard Fellowship, and the Sloan Fellowship.
\end{acknowledgments}

\newpage

\appendix

\section{Introduction to adjoint method}
In this section, we provide the general derivation of the adjoint method~\cite{Pontryagin2018,Chen2018a} and apply it to the master equation governing the dynamics of a dissipative quantum system.

\subsection{Derivation of the adjoint method}
Consider an initial value problem
\begin{equation}
    \frac{\text{d}\bm{v}}{\text{d}t} = f(\bm{v}, \bm{\theta})
\end{equation}
where the state $\bm{v}$ evolves from $t=0$ to $t=T$ and $\bm{\theta}$ are the time independent parameters for the ODE. Here we assume that there is a ``loss function'' $L(\bm{v}(T))$ which only depends on the final state $\bm{v}(T)$. This final state implicitly depends on both the initial state $\bm{v}(0)$ and the parameters $\bm \theta$ that specify its time evolution through the integration of the ODE. Our goal is to compute the gradient of $L$ with respect to both the initial state ($\partial L / \partial \bm{v}(0)$) and the ODE parameters ($\partial L / \partial \bm{\theta}$). Obtaining these gradients will enable optimization via gradient descent.

\subsubsection{Gradient with respect to the initial state}
The adjoint state is defined as $\bm{a}(t) = \partial L / \partial \bm{v}(t)$ and $\bm{a}(T)$ can be computed directly as long as $L$ is differentiable while $\bm{a}(0)$ is the gradient that we want. By the chain rule:
\begin{equation}
    \begin{split}
        a_i(t) =& \frac{\partial L}{\partial v_i(t)} = \sum_j \frac{\partial L}{\partial v_j(t+\epsilon)} \frac{\partial v_j(t+\epsilon)}{\partial v_i(t)} \\
        =& \sum_j a_j(t+\epsilon) \frac{\partial [v_j(t)+\epsilon f_j(\bm{v}(t), \bm{\theta})]}{\partial v_i(t)} \\
        =& a_i(t+\epsilon) + \epsilon \sum_j a_j(t+\epsilon) \frac{\partial  f_j(\bm{v}(t), \bm{\theta})}{\partial v_i(t)} .
    \end{split}
\end{equation}
Therefore the adjoint state satisfies the differential equation:
\begin{equation}
    \dot{a}_i(t) = \lim_{\epsilon \rightarrow 0} \frac{a_i(t+\epsilon)-a_i(t)}{\epsilon} = -\sum_j a_j(t) \frac{\partial f_j(\bm{v}(t), \bm{\theta})}{\partial v_i(t)}  .
\end{equation}
More compactly
\begin{equation}
    \dot{\bm{a}}(t) = -C^T \bm{a}(t) .
\end{equation}
where $\bm{a}$ is represented as a column vector and the matrix $C=\frac{\partial f(\bm{v}(t), \bm{\theta})}{\partial \bm{v}(t)}$ is defined as $C_{ij} = \frac{\partial  f_i(\bm{v}(t), \bm{\theta})}{\partial v_j(t)}$. In other words, solving the above differential equation from $t=T$ to $t=0$ allows us to obtain $\bm{a}(0) = \partial L / \partial \bm{v}(0)$.

\subsubsection{Gradient with respect to ODE parameters}
To compute the gradient $\partial L / \partial \bm{\theta}$, we embed the parameters $\bm\theta$ into the state vector and consider a new ODE:
\begin{equation}
    \frac{\text{d}}{\text{d}t} 
    \begin{pmatrix}
    \bm{v} \\
    \bm{\theta}
    \end{pmatrix}
    = 
    \begin{pmatrix}
    f(\bm{v}, \bm{\theta}) \\
    \bm{0}
    \end{pmatrix} .
\end{equation}
This leads to a composite adjoint satisfying
\begin{equation}
    \frac{\text{d}}{\text{d}t} 
    \begin{pmatrix}
    \bm{a}(t) \\
    \bm{a}_\theta (t)
    \end{pmatrix}
    = -
    \begin{pmatrix}
    \frac{\partial f(\bm{v}(t), \bm{\theta})}{\partial \bm{v}(t)} & \frac{\partial f(\bm{v}(t), \bm{\theta})}{\partial \bm{\theta}} \\
    \bm{0} & \bm{0}
    \end{pmatrix} ^T
    \begin{pmatrix}
    \bm{a}(t) \\
    \bm{a}_\theta (t)
    \end{pmatrix} .
\end{equation}
Define a matrix $D=\frac{\partial f(\bm{v}(t), \bm{\theta})}{\partial \bm{\theta}}$ as $D_{ij} = \frac{\partial  f_i(\bm{v}(t), \bm{\theta})}{\partial \theta_j(t)}$, then we have
\begin{equation}
    \dot{\bm{a}}_\theta (t) = - D^T \bm{a} (t) .
\end{equation}
Notice that $\bm{a}_\theta (T) = 0$ since $L$ doesn't explicitly depend on $\bm{\theta}$.

\subsection{Adjoint method for the master equation}
The above derivation works for general ODEs and here we would like to adapt the results to a specific type of ODE, the quantum master equation.

\subsubsection{Gradient with respect to the initial state}
It's easier to take derivatives with respect to the initial density matrix by working with superoperators.
Consider the linear transformation $\ket{i}\bra{j} \rightarrow \ket{i}\otimes\ket{j}$ that maps a density matrix $\rho = \sum \rho_{ij}\ket{i}\bra{j}$ to a state vector $\bar{\rho} = \sum \rho_{ij}\ket{i}\otimes\ket{j}$. By definition it's straightforward to check the following relation for left and right multiplication of an operator:
\begin{equation}
    A\rho B \rightarrow (A\otimes B^T) \bar{\rho} .
\end{equation}
Therefore the master equation Eq.~(\ref{eq:master}) is equivalent to
\begin{equation}
    \begin{split}
        \dot{\bar{\rho}} =& -i(H\otimes I - I\otimes H^T) \bar{\rho} + \sum_{k=1}^{K} \beta_k \left(A_k\otimes A_k^* \vphantom{\frac{1}{2}} \right. \\
        & \left. - \frac{1}{2} A_k^\dagger A_k \otimes I - \frac{1}{2} I \otimes (A_k^\dagger A_k)^T \right)\bar{\rho} \\
        =& M \bar{\rho}.
    \end{split}
\end{equation}
To derive the adjoint equation in presence of complex numbers, we could simply separate the real part (label $x$) and imaginary part (label $y$). The master equation can be written as
\begin{equation}\label{eq:super_real}
    \begin{split}
        & \frac{\text{d}}{\text{d} t}
        \begin{pmatrix}
        \bar{\rho}_x \\
        \bar{\rho}_y
        \end{pmatrix}
        = (M_x + iM_y)(\bar{\rho}_x + i\bar{\rho}_y) \\
        =&
        \begin{pmatrix}
        M_x \bar{\rho}_x - M_y \bar{\rho}_y \\
        M_x \bar{\rho}_y + M_y \bar{\rho}_x
        \end{pmatrix}
        = \begin{pmatrix}
        M_x & -M_y  \\
        M_y & M_x
        \end{pmatrix} 
        \begin{pmatrix}
        \bar{\rho}_x \\
        \bar{\rho}_y
        \end{pmatrix}\\
        =& \bm{M} 
        \begin{pmatrix}
        \bar{\rho}_x \\
        \bar{\rho}_y
        \end{pmatrix} .
    \end{split}
\end{equation}
Now we could calculate the $C$ matrix by
\begin{equation}
    C_{ij} = \frac{\partial [\bm{M}\rho(t)]_i}{\partial \rho_j(t)} = \frac{\partial \sum_{k}M_{ik}\rho_k(t)}{\partial \rho_j(t)} = M_{ij}
\end{equation}
which gives $C = \bm{M}$. Therefore the adjoint equation is given by
\begin{equation}
    \frac{\text{d}}{\text{d} t}
    \begin{pmatrix}
    \frac{\partial L}{\partial \bar{\rho}_x} \\
    \frac{\partial L}{\partial \bar{\rho}_y}
    \end{pmatrix}
    = - 
    \begin{pmatrix}
    M_x^T & M_y^T \\
    -M_y^T & M_x^T
    \end{pmatrix}
    \begin{pmatrix}
    \frac{\partial L}{\partial \bar{\rho}_x} \\
    \frac{\partial L}{\partial \bar{\rho}_y}
    \end{pmatrix} .
\end{equation}
This could be simplified by introducing the complex adjoint 
\begin{equation}
    \bar{a} = \frac{\partial L}{\partial \bar{\rho}_x} + i \frac{\partial L}{\partial \bar{\rho}_y} = \sum a_{ij} \ket{i}\otimes \ket{j}, \quad a_{ij} = \frac{\partial L}{\partial \rho_{ij}^x} + i \frac{\partial L}{\partial \rho_{ij}^y}
\end{equation}
such that
\begin{equation}
    \dot{\bar{a}} = -(M_x^T - i M_y^T) \left( \frac{\partial L}{\partial \bar{\rho}_x} + i \frac{\partial L}{\partial \bar{\rho}_y} \right) = -M^\dagger \bar{a} .
\end{equation}
More explicitly
\begin{equation}
    \begin{split}
        \dot{\bar{a}} =& -M^\dagger \bar{a} = -i(H\otimes I - I\otimes H^T) \bar{a} - \sum_{k=1}^{K} \beta_k \left(A_k^\dagger \otimes A_k^T \vphantom{\frac{1}{2}} \right. \\
        & \left. - \frac{1}{2} A_k^\dagger A_k \otimes I - \frac{1}{2} I \otimes (A_k^\dagger A_k)^T \right)\bar{a} .
    \end{split}
\end{equation}
Now we inverse the mapping and define the adjoint matrix
\begin{equation}
    \bar{a} = \sum a_{ij} \ket{i}\otimes \ket{j} \rightarrow a = \sum a_{ij} \ket{i} \bra{j}
\end{equation}
and the adjoint equation has a similar form as the original master equation:
\begin{equation}\label{eq:state_adjoint}
    \dot{a}=-i[H, a] - \sum_{k=1}^{K} \beta_k \left( A_k^\dagger a A_{k} - \frac{1}{2} \left \{ A_{k}^{\dagger}A_{k}, a \right\}\right) .
\end{equation}
Starting from $a(T)$ and solving the adjoint equation backward in time to $t=0$ gives $a(0)$, which is the gradient with respect to the initial density matrix.

\subsubsection{Gradient with respect to system parameters}
Going back to the superoperator representation Eq.~(\ref{eq:super_real}) and taking derivative with respect to some ODE parameter $\theta$:
\begin{equation}
    \begin{split}
        & \dot{a}_\theta (t) \\
        =& - \left[ \partial_\theta\bm{M} 
        \begin{pmatrix}
        \bar{\rho}_x \\
        \bar{\rho}_y
        \end{pmatrix} \right]^T 
        \begin{pmatrix}
        \frac{\partial L}{\partial \bar{\rho}_x} \\
        \frac{\partial L}{\partial \bar{\rho}_y}
        \end{pmatrix}
        = -(\bar{\rho}_x^T, \bar{\rho}_y^T) \partial_\theta\bm{M}^T
        \begin{pmatrix}
        \frac{\partial L}{\partial \bar{\rho}_x} \\
        \frac{\partial L}{\partial \bar{\rho}_y}
        \end{pmatrix} \\
        =& -(\bar{\rho}_x^T, \bar{\rho}_y^T)
        \begin{pmatrix}
        \partial_\theta M_x^T & \partial_\theta M_y^T \\
        -\partial_\theta M_y^T & \partial_\theta M_x^T
        \end{pmatrix} 
        \begin{pmatrix}
        \frac{\partial L}{\partial \bar{\rho}_x} \\
        \frac{\partial L}{\partial \bar{\rho}_y}
        \end{pmatrix} \\
        =& -(\bar{\rho}_x^T \partial_\theta M_x^T - \bar{\rho}_y^T \partial_\theta M_y^T, \bar{\rho}_x^T \partial_\theta M_y^T + \bar{\rho}_y^T \partial_\theta M_x^T)
        \begin{pmatrix}
        \frac{\partial L}{\partial \bar{\rho}_x} \\
        \frac{\partial L}{\partial \bar{\rho}_y}
        \end{pmatrix} \\
        =& - \left[ (\partial_\theta M \bar{\rho})_x^T \frac{\partial L}{\partial \bar{\rho}_x} + (\partial_\theta M \bar{\rho})_y^T \frac{\partial L}{\partial \bar{\rho}_y} \right] .
    \end{split}
\end{equation}
Define
\begin{equation}
    \bar{\Delta}_\theta = \partial_\theta M \bar{\rho} = \sum \Delta_{ij} \ket{i} \otimes \ket{j} \rightarrow \Delta_\theta = \sum \Delta_{ij} \ket{i} \bra{j}
\end{equation}
then the gradient could be simplified as
\begin{equation}\label{eq:adjoint_theta}
    \dot{a}_\theta (t) = -\sum_{ij}(\Delta_{ij}^x a_{ij}^x + \Delta_{ij}^y a_{ij}^y) = -\text{Re} \tr{\Delta_\theta a^\dagger(t)} .
\end{equation}
From the master equation, it's not hard to see that
\begin{equation}
    \begin{split}
        \Delta_{\alpha_j} =& -i[H_j, \rho(t)] \\
        \Delta_{\beta_k} =& A_k \rho(t) A_{k}^{\dagger} - \frac{1}{2} \left \{ A_{k}^{\dagger}A_{k}, \rho(t) \right\} .
    \end{split}
\end{equation}
Therefore after solving for $\rho(t)$ from Eq.~(\ref{eq:master}) and $a(t)$ from Eq.~(\ref{eq:state_adjoint}), we could solve $a_\theta(t)$ by integrating Eq.~(\ref{eq:adjoint_theta}) backward in time starting from $a_{\bm{\theta}}(T)=0$.

\section{Average fidelity}
We could evaluate the integration over Bloch sphere in the definition of average Fidelity (main text Eq.(\ref{eq:avg_F})) which leads to a simpler expression to work with
\begin{equation}\label{eq:F_avg_eval}
    \begin{split}
        \bar{F} (t) =& \tr{\left(\frac{1}{3} \hat{\rho}_{00} + \frac{1}{6} \hat{\rho}_{11}\right) \hat{\rho}_{00}(t)} \\
        &+ \tr{\left(\frac{1}{6} \hat{\rho}_{00} + \frac{1}{3} \hat{\rho}_{11}\right) \hat{\rho}_{11}(t)} \\
        &+ \text{Re} \left\{ \tr{\frac{1}{3} \hat{\rho}_{01} \hat{\rho}_{10}(t)} \right\} ,
    \end{split}
\end{equation}
For a single bosonic mode with $\hat{H}=0$ and only photon loss error, the average fidelity for subspace of $\ket{0}$ and $\ket{1}$ is
\begin{equation}
    \bar{F} (t) = \frac{1}{6} \left( e^{-\kappa t} + 2e^{-\kappa t /2} + 3 \right)
\end{equation}
where $\kappa$ is the loss rate. This is our definition of break-even fidelity throughout the paper.

Other fidelity definitions are also applicable as long as the gradients are computable. For example, we could learn AQEC with the entanglement fidelity~\cite{Albert2018}
\begin{equation}
    \bar{F}(t) = \frac{1}{4} \tr{\hat{\rho}_{00} \hat{\rho}_{00}(t) + \hat{\rho}_{11} \hat{\rho}_{11}(t) + \hat{\rho}_{01} \hat{\rho}_{10}(t) + \hat{\rho}_{10} \hat{\rho}_{01}(t)} .
\end{equation}
From our experience, we didn't observe any substantial differences in terms of training speed and optimization results when using entanglement fidelity compared to the average fidelity.

\subsection{Modified average fidelity}
In practice, we find out that there is a simple modification of the average fidelity that helps avoiding certain local minima during the training.
Instead of learning an identity map on the Bloch sphere, we could aim at preserving the Bloch sphere up to an arbitrary rotation in the logical subspace.
This modification extends the set of Hamiltonians that protects a given QEC code, which could potentially accelerate the \texttt{AutoQEC} searching.

Analytically deriving the modified average fidelity turns out to be challenging for general rotations on the Bloch sphere. We therefore restrict the allowed rotations to only along the $Z$ axis:
\begin{equation}
    \begin{split}
        \hat{U} \ket{\psi_0} =& \ket{\psi_0} \\
        \hat{U} \ket{\psi_1} =& e^{i\varphi} \ket{\psi_1} .
    \end{split}
\end{equation}
The modified average fidelity becomes
\begin{equation}\label{eq:modified_avg_F}
    \begin{split}
        \bar{F} (t) =& \max_{\hat{U}} \tr{\hat{U} \left(\frac{1}{3} \hat{\rho}_{00} + \frac{1}{6} \hat{\rho}_{11}\right) \hat{U}^\dagger \hat{\rho}_{00}(t)} \\
        &\qquad + \tr{\hat{U} \left(\frac{1}{6} \hat{\rho}_{00} + \frac{1}{3} \hat{\rho}_{11}\right) \hat{U}^\dagger \hat{\rho}_{11}(t)} \\
        &\qquad + \text{Re} \left\{ \tr{\frac{1}{3} \hat{U} \hat{\rho}_{01} \hat{U}^\dagger \hat{\rho}_{10}(t)} \right\} \\
        =& \max_{\varphi} \tr{\left(\frac{1}{3} \hat{\rho}_{00} + \frac{1}{6} \hat{\rho}_{11}\right) \hat{\rho}_{00}(t)} \\
        &\qquad + \tr{\left(\frac{1}{6} \hat{\rho}_{00} + \frac{1}{3} \hat{\rho}_{11}\right) \hat{\rho}_{11}(t)} \\
        &\qquad + \text{Re} \left\{ \tr{\frac{1}{3} e^{-i\varphi} \hat{\rho}_{01} \hat{\rho}_{10}(t)} \right\} \\
        =& \tr{\left(\frac{1}{3} \hat{\rho}_{00} + \frac{1}{6} \hat{\rho}_{11}\right) \hat{\rho}_{00}(t)} \\
        &+ \tr{\left(\frac{1}{6} \hat{\rho}_{00} + \frac{1}{3} \hat{\rho}_{11}\right) \hat{\rho}_{11}(t)} \\
        &+ \left | \tr{\frac{1}{3} \hat{\rho}_{01} \hat{\rho}_{10}(t)} \right | ,
    \end{split}
\end{equation}
where the only change compared to Eq.~(\ref{eq:F_avg_eval}) is to replace the real part of $\tr{\hat{\rho}_{01} \hat{\rho}_{10}(t)}$ with its absolute value.
Throughout this paper, we use this modified average fidelity (Eq.~(\ref{eq:modified_avg_F})) as the objective function for \texttt{AutoQEC}.

\section{Derivation of the $\sqrt{3}$ code}\label{appendix:b}
With $d=2$ Hamiltonian, \texttt{AutoQEC} discovered an error correcting code which we were not able to find in literature. This simple ``$\sqrt{3}$" code warrants further investigation, and here we derive it analytically based on the Hamiltonian distance constraints and QEC properties. Notice that the AQEC Hamiltonian for a given code is not unique -- we therefore make some assumptions about the Hamiltonian structure to simplify the derivation.

\subsection{General results}
Consider the problem of correcting a single photon loss error with a bosonic mode. The logical states are $\ket{\psi_0}$ and $\ket{\psi_1}$. For simplicity, we assume the error states $\ket{\psi_2} \propto \hat a\ket{\psi_0}$ and $\ket{\psi_3} \propto \hat a\ket{\psi_1}$ are also mutually orthogonal to both logical states. Therefore $\{\ket{\psi_0}, \ket{\psi_1}, \ket{\psi_2}, \ket{\psi_3}\}$ forms the basis for a 4-dimensional subspace $\mathcal{H}_1$. We choose the Hilbert space cutoff $\ket{N}$ as the highest Fock level that has a non-zero overlap with either $\ket{\psi_0}$ or $\ket{\psi_1}$ (total Hilbert space dimension $N+1$). Now the Hilbert space is decomposed into $\mathcal{H}_1$ and its orthogonal complement $\mathcal{H}_2$ ($\mathcal{H} = \mathcal{H}_1 \oplus \mathcal{H}_2$) where a set of orthogonal basis for $\mathcal{H}_2$ is $\{\ket{\psi_4},...,\ket{\psi_N}\}$.

To correct the single photon loss error autonomously, the Hamiltonian should include the following terms
\begin{equation}\label{eq:QEC_1}
    \hat H = (\ket{\psi_0}\bra{\psi_2} + \ket{\psi_1}\bra{\psi_3}) \otimes \ket{e}\bra{g} + \mathrm{h.c.}
\end{equation}
which basically maps the error states to the correct logical states and excited the qubit: $\ket{\psi_2,g} \leftrightarrow \ket{\psi_0,e}$ and $\ket{\psi_3,g} \leftrightarrow \ket{\psi_1,e}$. After relaxation of the ancilla qubit, the error states $\ket{\psi_2}$ and $\ket{\psi_3}$ are mapped back to the logical states $\ket{\psi_0}$ and $\ket{\psi_1}$ while maintaining the relative phase between them due to the identical Rabi rate in the Hamiltonian.
Adiabatically eliminating the ancilla qubit results in an effective dissipator $D[\ket{\psi_0}\bra{\psi_2} + \ket{\psi_1}\bra{\psi_3}]$, which provides an alternative way of understanding the reduced dynamics for the bosonic mode.

The Hamiltonian Eq.~(\ref{eq:QEC_1}) is conceptually simple, but may be difficult to generate in experiment since it can be highly non-local in the Fock basis. In the case of the smallest binomial code $\ket{\psi_0} = \frac{1}{\sqrt{2}} (\ket{0} + \ket{4})$ and $\ket{\psi_1} = \ket{2}$, realizing Eq.~(\ref{eq:QEC_1}) would require a coupling term $\ket{3,g}\bra{0,e}$ which does not occur naturally~\cite{Hu2019,Gertler2021}. Going to a higher order binomial code is even worse since it requires even more non-local interaction between Fock states.

This is where states in the orthogonal subspace $\mathcal{H}_2$ could contribute. The basic idea is that those states won't change the QEC behavior, but their proper combinations could cancel certain non-local interactions in Eq.~(\ref{eq:QEC_1}) and make the total Hamiltonian easier to implement. For this purpose, the general form of the QEC Hamiltonian is (we ignore all $\ket{g}\bra{g}$ and $\ket{e}\bra{e}$ terms since those won't solve the locality problem anyway)
\begin{equation}\label{eq:QEC_full}
    \begin{split}
        \hat{H} =& \tilde{H}^\dagger \otimes \ket{e}\bra{g} + \tilde{H} \otimes \ket{g}\bra{e} \\
        \tilde{H} =& \ket{\psi_2}\bra{\psi_0} + \ket{\psi_3}\bra{\psi_1} + \sum_{i,j}^{N} \beta_{ij} \ket{\psi_i} \bra{\psi_j}.
    \end{split}
\end{equation}
We impose a number of constraints on the coefficients $\beta_{ij}$ such that the summation part of $\tilde{H}$ does not modify the error correction dynamics. In particular, we require that $\beta_{ij}=0$ for $i<4$ and $j<2$. The requirement $j<2$ removes overlap with the states $\ket{\psi_{0\sim 1},e}$ -- we want only the first part of $\tilde{H}$ to perform this correction function. Similarly, removing $i<4$ prevents overlap with states $\ket{\psi_{0\sim 3},g}$ which is important for preventing the summation part of Hamiltonian from causing states to transition into the code and error subspaces.

Having set these conditions on $\tilde{H}$, we now quantify the notion of locality for the Hamiltonian. We define the distance of a Hamiltonian $H$ to be $d$ if $\tilde{H}_{mn}\equiv\mele{m}{\tilde{H}}{n} = 0, \forall |m-n|>d, 0\leq m,n \leq N$. From our numerical search (Fig.~\ref{fig2}), we found that $d=1$ Hamiltonians generate only trivial error correcting codes. Therefore we will consider the $d=2$ case, which seems to be the minimal distance required for QEC exceeding break-even. More explicitly, a $d=2$ Hamiltonian satisfies
\begin{equation}\label{eq:locality_constraints}
    \begin{pmatrix}
    0 & \tilde{H}_{03} & \tilde{H}_{04} & ... & \tilde{H}_{0N} \\
    \tilde{H}_{30} & 0 & \tilde{H}_{14} & ... & \tilde{H}_{1N} \\
    \tilde{H}_{40} & \tilde{H}_{41} & 0 & ... & \tilde{H}_{2N} \\
    \vdots & & & & \tilde{H}_{N-3,N} \\
    \tilde{H}_{N0} & \tilde{H}_{N1} & ... & \tilde{H}_{N,N-3} & 0
    \end{pmatrix} =\mathbf{0}.
\end{equation}
The goal here is to find $\tilde{H}$, in other words we will solve for the coefficients $\beta_{ij}$ as well as the logical states while satisfying these locality constraints.

\subsection{Example of the $\sqrt{3}$ code}
We demonstrate how to solve this problem with a concrete example. Numerically the $\sqrt{3}$ code we found with the $d=2$ Hamiltonian had logical states of the form
\begin{equation}
    \begin{split}
        \ket{\psi_0} &= a_0 \ket{0} + a_3 \ket{3} \\
        \ket{\psi_1} &= a_1 \ket{1} + a_4 \ket{4} + a_6 \ket{6} .
    \end{split} 
\end{equation}
Since the two logical states don't share any Fock basis, we can always make all coefficients $a_0,a_3,a_1,a_4,a_6$ real by doing the basis transformation $\ket{n} \rightarrow e^{i\theta_n} \ket{n}$. The error states are
\begin{equation}
    \begin{split}
        \ket{\psi_2} &= \ket{2} \propto \hat a\ket{\psi_0} \\
        \ket{\psi_3} &= \mathcal{N}_1 ( a_1 \ket{0} + 2 a_4 \ket{3} + \sqrt{6} a_6 \ket{5} ) \propto \hat a\ket{\psi_1} .
    \end{split} 
\end{equation}
Notice that if Eq.~(\ref{eq:locality_constraints}) does have a solution, the solution always exists no matter how we choose the basis for the orthogonal subspace $\mathcal{H}_2$. In other words, we could always represent the new basis as linear combinations of the old basis and that together with the old solution $\beta_{ij}$ gives the new solution $\beta_{ij}'$. Therefore here we have complete freedom to select the basis $\{ \ket{\psi_4},\ket{\psi_5},\ket{\psi_6} \}$ for $\mathcal{H}_2$ and for convenience of further analysis we make the following choice (notation $\psi_i(n) = \inp{n}{\psi_i}$):
\begin{equation}
    \begin{split}
        \ket{\psi_4} & = \psi_4(1) \ket{1} + \psi_4(4) \ket{4} + \psi_4(6) \ket{6} \\
        \ket{\psi_5} & = \psi_5(1) \ket{1} + \psi_5(4) \ket{4} + \psi_5(6) \ket{6} \\
        \ket{\psi_6} & = \psi_6(0) \ket{0} + \psi_6(3) \ket{3} + \psi_6(5) \ket{5} .
    \end{split} 
\end{equation}
We can make all $\psi_i(n)$ to be real here, which leads to all $\beta_{ij}$ also being real.
With this basis choice, many constraints in Eq.~(\ref{eq:locality_constraints}) can be easily satisfied either automatically or by setting certain $\beta_{ij}=0$. More specifically, for any $|m-n|>2$ such that $\mele{m}{(\ket{\psi_2}\bra{\psi_0} + \ket{\psi_3}\bra{\psi_1})}{n} = 0$, there are two different cases:
\begin{enumerate}
    \item $\mele{m}{(\ket{\psi_i}\bra{\psi_j})}{n}=0, \forall i,j$: in this case $\tilde{H}_{mn}=0$ is already satisfied.
    \item there exists $i,j$ such that $\mele{m}{(\ket{\psi_i}\bra{\psi_j})}{n}\neq 0$: in this case we just set $\beta_{ij}=0$.
\end{enumerate}
Therefore the only non-trivial constraints from Eq.~(\ref{eq:locality_constraints}) are those with $\mele{m}{(\ket{\psi_2}\bra{\psi_0} + \ket{\psi_3}\bra{\psi_1})}{n} \neq 0$ which are $\tilde{H}_{04},\tilde{H}_{06},\tilde{H}_{36},\tilde{H}_{51}$. It is easy to see that the only terms in Eq.~(\ref{eq:QEC_full}) that will contribute to these matrix elements are $\ket{\psi_6}\bra{\psi_4}$ and $\ket{\psi_6}\bra{\psi_5}$. With these analysis, the ansatz Hamiltonian Eq.~(\ref{eq:QEC_full}) can greatly simplified to the following:
\begin{equation}\label{eq:H1}
    \tilde{H} = \ket{\psi_2}\bra{\psi_0} + \ket{\psi_3}\bra{\psi_1} + \beta_1 \ket{\psi_6} \bra{\psi_4} + \beta_2 \ket{\psi_6} \bra{\psi_5} ,
\end{equation}
where the two free parameters $\beta_1$ and $\beta_2$ satisfy a set of linear equations
\begin{subequations}
    \begin{align}
        \tilde{H}_{04} &= \psi_3(0) \psi_1(4) + \beta_1 \psi_6(0) \psi_4(4) + \beta_2 \psi_6(0) \psi_5(4) = 0 \label{eq:a} \\
        \tilde{H}_{06} &= \psi_3(0) \psi_1(6) + \beta_1 \psi_6(0) \psi_4(6) + \beta_2 \psi_6(0) \psi_5(6) = 0 \label{eq:b} \\
        \tilde{H}_{36} &= \psi_3(3) \psi_1(6) + \beta_1 \psi_6(3) \psi_4(6) + \beta_2 \psi_6(3) \psi_5(6)  = 0 \label{eq:c} \\
        \tilde{H}_{51} &= \psi_3(5) \psi_1(1) + \beta_1 \psi_6(5) \psi_4(1) + \beta_2 \psi_6(5) \psi_5(1) = 0 \label{eq:d} .
    \end{align}
\end{subequations}
The crucial observation here is that the number of equations 4 is larger than the number of parameters 2, which means the coefficients must be linearly dependent. Since these coefficients are essentially functions of $\ket{\psi_0}$ and $\ket{\psi_1}$, this eventually provides the extra constraints for determining the logical states. Here there should be $4-2=2$ constraints in total.

Below we will show in details how to obtain the two constraints and eventually the two logical states. Comparing Eq.~(\ref{eq:b}) and Eq.~(\ref{eq:c}), it's easy to see that the first constraint is
\begin{equation}\label{eq:constraint1}
    \frac{\psi_3(0)}{\psi_3(3)} = \frac{\psi_6(0)}{\psi_6(3)} .
\end{equation}
To get the second constraint, let us multiply Eq.~(\ref{eq:a}) with $\psi_1(4)$, multiply Eq.~(\ref{eq:b}) with $\psi_1(6)$, and then add them together:
\begin{equation}
    \begin{split}
        & \psi_3(0) ( [\psi_1(4)]^2 + [\psi_1(6)]^2 )  \\
        & +\beta_1 \psi_6(0) [ \psi_4(4) \psi_1(4) + \psi_4(6) \psi_1(6) ] \\
        & + \beta_2 \psi_6(0) [ \psi_5(4) \psi_1(4) + \psi_5(6) \psi_1(6) ] = 0 .
    \end{split}
\end{equation}
Using the fact that $\ket{\psi_1}$ is normalized and orthogonal to both $\ket{\psi_4}$ and $\ket{\psi_5}$, we have
\begin{equation}
    \begin{split}
        & \psi_3(0) ( 1 - [\psi_1(1)]^2) + \beta_1 \psi_6(0) [ -\psi_4(1) \psi_1(1) ] \\
        & + \beta_2 \psi_6(0) [ -\psi_5(1) \psi_1(1) ] = 0 \\
        \Rightarrow &  - \psi_3(0) \frac{1 - [\psi_1(1)]^2}{\psi_1(1)} + \beta_1 \psi_6(0) \psi_4(1) \\
        & + \beta_2 \psi_6(0) \psi_5(1) = 0 .
    \end{split}
\end{equation}
Compare this with Eq.~(\ref{eq:d}), we immediately obtain the second constraint:
\begin{equation}\label{eq:constraint2}
    \begin{split}
        & - \psi_3(0) \frac{1 - [\psi_1(1)]^2}{\psi_1(1) \psi_6(0)} = \frac{\psi_3(5) \psi_1(1)}{\psi_6(5)} \\
        \Rightarrow & \psi_3(0) \psi_6(5) (1 - [\psi_1(1)]^2) + \psi_3(5) \psi_6(0) [\psi_1(1)]^2 = 0 .
    \end{split}
\end{equation}
Let us explicitly list all the relevant states here
\begin{equation}
    \begin{split}
        \ket{\psi_0} &= a_0 \ket{0} + a_3 \ket{3} \\
        \ket{\psi_1} &= a_1 \ket{1} + a_4 \ket{4} + a_6 \ket{6} \\
        \ket{\psi_3} &= \mathcal{N}_1 ( a_1 \ket{0} + 2 a_4 \ket{3} + \sqrt{6} a_6 \ket{5} ) \\
        \ket{\psi_6} &= \mathcal{N}_2 ( a_1 \ket{0} + 2 a_4 \ket{3} + \beta \ket{5} ) ,
    \end{split} 
\end{equation}
where we have applied Eq.~(\ref{eq:constraint1}) for $\ket{\psi_6}$ and $\beta$ is another parameter. Combining the QEC criteria and Eq.~(\ref{eq:constraint2}), we have
\begin{equation}
    \begin{split}
        & a_0^2 + a_3^2 = 1 \\
        & a_1^2 + a_4^2 + a_6^2 = 1 \\
        & a_0 a_1 + 2 a_3 a_4 = 0 \\
        & 3 a_3^2 = a_1^2 + 4 a_4^2 + 6 a_6^2 \\
        & a_1^2 + 4 a_4^2 + \sqrt{6}\beta a_6 = 0 \\
        & \beta (1 - a_1^2) + \sqrt{6} a_6 a_1^2 = 0 .
    \end{split}
\end{equation}
We have 6 equations and 6 parameters in total, and the solution is (there is some freedom to choose the signs which again is just a trivial basis transformation)
\begin{equation}
    \begin{split}
        & a_0 = \sqrt{1-\frac{1}{\sqrt{3}}}, a_3 = \frac{1}{\sqrt[4]{3}}, a_1 = \sqrt{\frac{2(6-\sqrt{3})}{\sqrt{3}+9}} \\
        & a_4 = -\sqrt{\frac{(\sqrt{3}-1)(6-\sqrt{3})}{2(\sqrt{3}+9)}}, a_6 = \sqrt{\frac{3-\sqrt{3}}{2(\sqrt{3}+9)}} .
    \end{split}
\end{equation}
Therefore the logical states of the $\sqrt{3}$ code are
\begin{equation}\label{eq:logical_states}
    \begin{split}
        \ket{\psi_0} =& \sqrt{1-\frac{1}{\sqrt{3}}} \ket{0} + \frac{1}{\sqrt[4]{3}} \ket{3} \\
        \ket{\psi_1} =& \sqrt{\frac{2(6-\sqrt{3})}{\sqrt{3}+9}} \ket{1} - \sqrt{\frac{(\sqrt{3}-1)(6-\sqrt{3})}{2(\sqrt{3}+9)}} \ket{4} \\
        & + \sqrt{\frac{3-\sqrt{3}}{2(\sqrt{3}+9)}} \ket{6} .
    \end{split}
\end{equation}
Notice that the average number of photons in the codewords is $3|a_3|^2 = \sqrt{3}$.

Now we could complete all basis states and the Hamiltonian Eq.~(\ref{eq:H1}). The basis of $\mathcal{H}_2$:
\begin{equation}
    \begin{split}
        \ket{\psi_6} &= \mathcal{N}_2 ( a_1 \ket{0} + 2 a_4 \ket{3} + \beta \ket{5} ) , \quad \beta = -\frac{a_1^2+4a_4^2}{\sqrt{6} a_6} \\
        \ket{\psi_4} &= \mathcal{N}_3 ( a_4 \ket{1} - a_1 \ket{4} ) \\
        \ket{\psi_5} &= \mathcal{N}_4 ( a_1 \ket{1} + a_4 \ket{4} + \beta' \ket{6} ) , \quad \beta' = -\frac{a_1^2+a_4^2}{a_6} 
    \end{split} 
\end{equation}
and the Hamiltonian parameters:
\begin{equation}
    \beta_2 = -\frac{\mathcal{N}_1 a_6}{\mathcal{N}_2 \mathcal{N}_4 \beta'} \qquad \beta_1 = \frac{\mathcal{N}_1 a_4 (1-a_6/\beta')}{\mathcal{N}_2 \mathcal{N}_3 a_1} .
\end{equation}

There are some extra complexities in constructing the AQEC Hamiltonian and we actually need to keep more terms from the summation in Eq.~(\ref{eq:QEC_full}) rather than just the $\beta_1$ and $\beta_2$ terms in Eq.~(\ref{eq:H1}). To understand why this is required, let us study a simpler problem of stabilizing $\ket{\psi} = \frac{1}{\sqrt{2}}(\ket{0}+\ket{2})$ under photon loss error.
Even though the Hamiltonian $\hat{H} = (\ket{0,e}+\ket{2,e})\bra{1,g} + \text{h.c.}$ corrects the error after a single photon loss, it doesn't actually lead to state stabilization.
The reason is that when no photon loss happens, the state evolves within the subspace $\{\ket{0}, \ket{2}\}$ under the non-Hermitian Hamiltonian $\hat{H}'=-i\kappa \hat{a}^\dagger \hat{a}/2$ and eventually becomes $\ket{0}$. This is because non-detection of a photon still provides us information about the state causing us to update it in a way that skews towards a lower number of photons. State-stabilization must undo this effect.
We protect $\ket{\psi}$ against $\hat{H}'$ by engineering a large detuning for $\ket{\psi}$ within the subspace $\{\ket{0}, \ket{2}\}$. For example, adding extra terms such as $\Omega \ket{\psi} \bra{\psi}$ or $\Omega \ket{0}\bra{2}+\text{h.c.}$ to $\hat{H}$ will stabilize $\ket{\psi}$. This new interaction can be seen as rapidly repopulating the $\ket{2}$ component of the wavevector as it decays through non-detection of photons.

Similarly, Eq.~(\ref{eq:H1}) only protects the logical states against single photon loss error, but not the non-unitary dynamics under $\hat{H}'$.
Fortunately, keeping a few extra terms from the summation in Eq.~(\ref{eq:QEC_full}) is sufficient to generate the large detuning without changing the Hamiltonian distance as well as the above derivation.
The choices are not unique and one option is to add $\ket{\psi_4} \bra{\psi_2}$ as well as $(a_6 \ket{4} - a_4 \ket{6}) \bra{5}$ in $\tilde{H}$, which produces similar results compared to the discovered code in Fig.~\ref{fig2}(c)i.
On the other hand, all these complications in constructing a proper AQEC Hamiltonian are automatically taken care of by \texttt{AutoQEC} through  numerical optimization of the average fidelity.


\section{Minimizing the emission bandwidth $\mathcal{B}$}
Here we prove the claim in the main text that for a harmonic oscillator coupled to a three-level qubit with Hamiltonian $\hat{H}_{ab}$ in Eq.~(\ref{eq:H_ab}), the emission bandwidth $\mathcal{B}$ is minimized when $g_1^2/\Delta_1 \approx g_2^2/\Delta_2$.

\begin{proof}
The Hamiltonian can be written in the subspace of $\{\ket{n+2,g},\ket{n+1,e},\ket{n,f}\}$ as a matrix
\begin{equation}
    \begin{pmatrix}
    0 & \sqrt{n+2} g_1 & 0 \\
    \sqrt{n+2} g_1 & \Delta_1 & \sqrt{n+1} g_2 \\
    0 & \sqrt{n+1} g_2 & \Delta_2
    \end{pmatrix} 
\end{equation}
and the eigenvalues satisfy
\begin{equation}
    \begin{split}
        & \lambda^3 - (\Delta_1+\Delta_2) \lambda^2 + \left[\Delta_1 \Delta_2 - (n+2) g_1^2 - (n+1) g_2^2 \right] \lambda \\
        & + (n+2) g_1^2 \Delta_2 = 0 .
    \end{split}
\end{equation}
In the dispersive regime $\Delta_{1,2} \gg g_{1,2}$, the eigenvalues can be expanded perturbatively as
\begin{equation}
    \begin{split}
        \lambda =& \lambda_0 + \lambda_1 + \lambda_2 + \mathcal{O}\left(\frac{g^3}{\Delta^3}\right) g \\
        \lambda_1 =& \mathcal{O}\left(\frac{g}{\Delta}\right) g, \quad \lambda_2 = \mathcal{O}\left(\frac{g^2}{\Delta^2}\right) g .
    \end{split}
\end{equation}
For dressed eigenstates $\ket{\widetilde{n+2,g}}$, $\lambda_0 = 0$ and
\begin{equation}
    \begin{split}
        & \left[\Delta_1 \Delta_2 - (n+2) g_1^2 - (n+1) g_2^2 \right] \lambda_1 + (n+2) g_1^2 \Delta_2 = 0 \\ 
        \Rightarrow & \quad \lambda_1 = -\frac{(n+2) g_1^2}{\Delta_1}
    \end{split}
\end{equation}
which agrees with the dispersive coupling Hamiltonian and no level nonlinearity shows up at this order. To the next order,
\begin{equation}
    \begin{split}
        & \left[\Delta_1 \Delta_2 - (n+2) g_1^2 - (n+1) g_2^2 \right] (\lambda_1 + \lambda_2) \\
        & - (\Delta_1+\Delta_2) \lambda_1^2  + (n+2) g_1^2 \Delta_2 = 0
    \end{split}
\end{equation}
which gives
\begin{equation}
    \lambda_2 = \frac{(n+2)g_1^2}{\Delta_1^2} \left[ (n+2) \frac{g_1^2}{\Delta_1} - (n+1) \frac{g_2^2}{\Delta_2} \right] .
\end{equation}
Notice that in general $\lambda_2$ will induce nonlinearity for the dressed states $\ket{\widetilde{n,g}}$ since it depends on $n^2$. However, when $g_1^2/\Delta_1 = g_2^2/\Delta_2$ the dependence on $n^2$ is completely removed which means the nonlinearity and therefore also the emission bandwidth $\mathcal{B}$ is eliminated at this order.
\end{proof}

\subsection{Qubit choice for $\hat{b}$}
The relevant dispersive coupling to the $e$ levels is
\begin{equation}
    \chi_e = \frac{2g_1^2}{\Delta_1} - \frac{g_2^2}{\Delta_2 - \Delta_1}
\end{equation}
and at the minimal nonlinearity point, we have
\begin{equation}
    \chi_e = \frac{g_1^2}{\Delta_1} \frac{r-2}{r-1}
\end{equation}
where $r=g_2^2/g_1^2=\Delta_2/\Delta_1$.
Ideally, $\chi_e$ should be as large as possible at this minimal nonlinearity point, such that we can selectively drive certain level transitions without introducing large $\mathcal{B}$.
For a transmon qubit $r \approx 2 \Rightarrow \chi_e \approx 0$ and therefore cannot be used as qubit $\hat{b}$.
Fortunately, other qubit designs could provide much more flexibility in engineering the coupling ratio $r$ and $r \approx 1$ is favorable in terms of larger $\chi_e$.

In this work, we choose a fluxonium type of Hamiltonian
\begin{equation}
    \hat{H} = 4 E_C \hat{n}^2 - E_J \cos (\hat{\phi} - \phi_{\text{ext}}) + \frac{1}{2} E_L \hat{\phi}^2
\end{equation}
for qubit $\hat{b}$. With realistic parameters $\phi_{\text{ext}}=0$, $E_C/2\pi=0.95~\text{GHz}$, $E_J/2\pi=4.75~\text{GHz}$ and $E_L/2\pi=0.65~\text{GHz}$, the coupling ratio is $r =  g_2^2/g_1^2 = |\mele{f}{\hat{n}}{\hat{e}}|^2/|\mele{e}{\hat{n}}{\hat{g}}|^2 \approx 1.2$ with $\omega_{ge}/2\pi \approx 5.43~\text{GHz}$ and $\omega_{ef}/2\pi \approx 3.87~\text{GHz}$.

\section{Full circuit design}
In this section, we provide details for the full circuit simulation in Fig.~\ref{fig4}(e).
The AQEC Hamiltonian Eq.~(\ref{eq:AQEC_H}) can be implemented with a more physical Hamiltonian
\begin{equation}\label{eq:H_physical}
    \hat{H} = \hat{H}_{ab} + \left( f_1(t)\hat{a}^\dagger + f_2(t) \hat{a} + f_3(t) \hat{a}^2 \right) \hat{b}^\dagger + f_4(t) \hat{b}^\dagger \hat{c} + \text{h.c.} 
\end{equation}
where
\begin{equation}
    \begin{split}
        f_1(t) =& \sum_n \frac{\alpha^{(1)}_n e^{-i(E_{n,e} - E_{n-1,g})t}}{\mele{\widetilde{n,e}}{\hat{a}^\dagger \hat{b}^\dagger}{\widetilde{n-1,g}}} \\
        f_2(t) =& \sum_n \frac{\alpha^{(2)}_n e^{-i(E_{n,e} - E_{n+1,g})t}}{\mele{\widetilde{n,e}}{\hat{a} \hat{b}^\dagger}{\widetilde{n+1,g}}} \\
        f_3(t) =& \sum_n \frac{\alpha^{(3)}_n e^{-i(E_{n,e} - E_{n+2,g})t}}{\mele{\widetilde{n,e}}{\hat{a}^2 \hat{b}^\dagger}{\widetilde{n+2,g}}} \\
        f_4(t) =& \Omega \sum_n \frac{e^{-i(E_{n,e} - E_{n,g})t}}{\mele{\widetilde{n,e}}{\hat{b}^\dagger}{\widetilde{n,g}}} 
    \end{split}
\end{equation}
and $\ket{\widetilde{n,g(e)}}$ are the dressed eigenstates of $\hat{H}_{ab}$ with energies $E_{n,g(e)}$.
Notice that the dressed states $\ket{\widetilde{n,g}}$ replace the bare Fock states $\ket{n,g}$ in our definition of the logical basis in Eq.~(\ref{eq:logical_states}) and the matrix elements such as $\mele{\widetilde{n,e}}{\hat{a}^\dagger \hat{b}^\dagger}{\widetilde{n-1,g}}$ will be close but not equal to $\sqrt{n}$.
The drivings $f_{1\sim 4}(t)$ engineer couplings between dressed states rather than the bare Fock states.

The relevant dissipators are $\{ \sqrt{\kappa} \hat{a}, \sqrt{\kappa_q} \hat{c} \}$, but to be more realistic we also include an extra dissipator $\sqrt{\kappa} \hat{b}$ in the simulation.
The coupling strength $\Omega$ between $\hat{b}$ and $\hat{c}$ is chosen such that the effective decay rate for $\hat{b}$ after adiabatically eliminating $\hat{c}$~\cite{Reiter2012a} is still $4\Omega^2/\kappa_q=2\pi \times 20~\text{MHz}$, same as the value used in the numerical optimization. We set the decay rate of $\hat{c}$ as $\kappa_q/2\pi=100~\text{MHz}$.

The Hamiltonian Eq.~(\ref{eq:H_physical}) can be furthermore implemented with a circuit model
\begin{equation}\label{eq:H_circuit}
    \begin{split}
        \hat{H} =& \omega_a \hat{a}^\dagger \hat{a} + \omega_{ge} \ket{e} \bra{e} + (\omega_{ge} + \omega_{ef}) \ket{f} \bra{f}+ \omega_c \hat{c}^\dagger \hat{c}  \\
        &+ g_1 (\hat{a}^\dagger \ket{g}\bra{e} + \hat{a}\ket{e}\bra{g}) + g_2 (\hat{a}\ket{f}\bra{e} + \hat{a}^\dagger \ket{e}\bra{f}) \\
        &+ \varepsilon_1 (t) \left[ g_{ab}^{(1)} \cos \left( \varphi_a (\hat{a}+\hat{a}^\dagger) + \varphi_b (\hat{b}+\hat{b}^\dagger) \right) \right. \\
        &\qquad \qquad \left. + g_{bc}^{(1)} \cos \left( \varphi_b (\hat{b}+\hat{b}^\dagger) + \varphi_c (\hat{c}+\hat{c}^\dagger) \right) \right] \\
        &+ \varepsilon_2 (t) \left[ g_{ab}^{(2)} \sin \left( \varphi_a (\hat{a}+\hat{a}^\dagger) + \varphi_b (\hat{b}+\hat{b}^\dagger) \right) \right. \\
        &\qquad \qquad \left. + g_{bc}^{(2)} \sin \left( \varphi_b (\hat{b}+\hat{b}^\dagger) + \varphi_c (\hat{c}+\hat{c}^\dagger) \right) \right] ,
    \end{split}
\end{equation}
where the drivings are given by
\begin{equation}
    \begin{split}
        \varepsilon_1 (t) =& - 2 \text{Re} \left\{ \frac{1}{\varphi_a \varphi_b g_{ab}^{(1)}} \left[ e^{-2i\omega_a t} f_1(t) + f_2(t)  \right] \right. \\
        &\left. + \frac{1}{\varphi_b \varphi_c g_{bc}^{(1)}} e^{i(\omega_c- \omega_a) t} f_4(t) \right\} \\
        \varepsilon_2 (t) =& - 2 \text{Re} \left\{ \frac{2}{\varphi_a^2 \varphi_b g_{ab}^{(2)}} e^{i \omega_a t} f_3(t) \right\} ,
    \end{split}
\end{equation}
which are generated by the two independent flux pump through the larger and smaller loops~\cite{Kapit2016a} in Fig.~\ref{fig4}(c).
After Taylor expanding the $\cos$ and $\sin$ interaction and dropping fast rotating terms in the rotating frame, we could show the equivalence of Eq.~(\ref{eq:H_circuit}) to the Hamiltonian Eq.~(\ref{eq:H_physical}) with $\Delta_1 = \omega_{ge}-\omega_a$ and $\Delta_2 = \Delta_1 + \omega_{ef}-\omega_a$.
To ensure the validity of rotating wave approximation, we place the frequencies at $\omega_a/2\pi=3.5~\text{GHz}$ and $\omega_c/2\pi=2.5~\text{GHz}$ with qubit $\hat{b}$ frequencies from the previous section. We also choose $\varphi_a=\varphi_b=\varphi_c=0.1$ such that higher order terms in the $\cos$ and $\sin$ expansions can be safely dropped.
All AQEC Hamiltonian parameters as well as the logical basis states are directly imported from the \texttt{AutoQEC} optimization result instead of using the analytical results in Appendix~\ref{appendix:b}.
We use QuTiP~\cite{Johansson2012,Johansson2013} for the full circuit simulation.

\begin{figure}[t]
    \centering
    \includegraphics[width=0.4\textwidth]{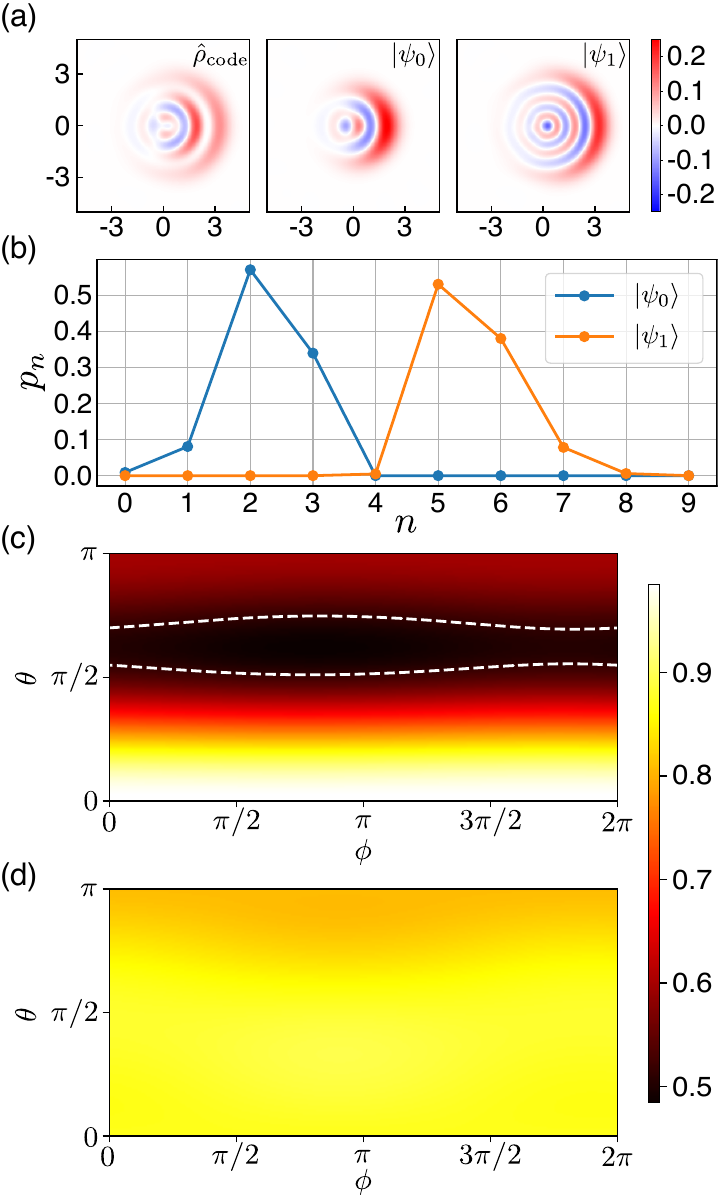}
    \caption{(a-b) Wigner functions and photon number distributions for the discovered encoding in Fig.~\ref{fig2}(c)ii. (c) Single state fidelity $F_{\theta\phi}$ at $t=10\mu$s for the whole Bloch sphere. The white dashed line indicates the break-even fidelity. (d) $F_{\theta\phi}$ on the Bloch sphere for the $\sqrt{3}$ code. For all Wigner funciton plots throughout this paper, the horizontal axis label is $x = \ave{\hat{a} + \hat{a}^\dagger}/\sqrt{2}$ and the vertical axis label is $p = i\ave{\hat{a}^\dagger - \hat{a}}/\sqrt{2}$.}
    \label{fig_SI_1}
\end{figure}

\begin{figure}[t]
    \centering
    \includegraphics[width=0.4\textwidth]{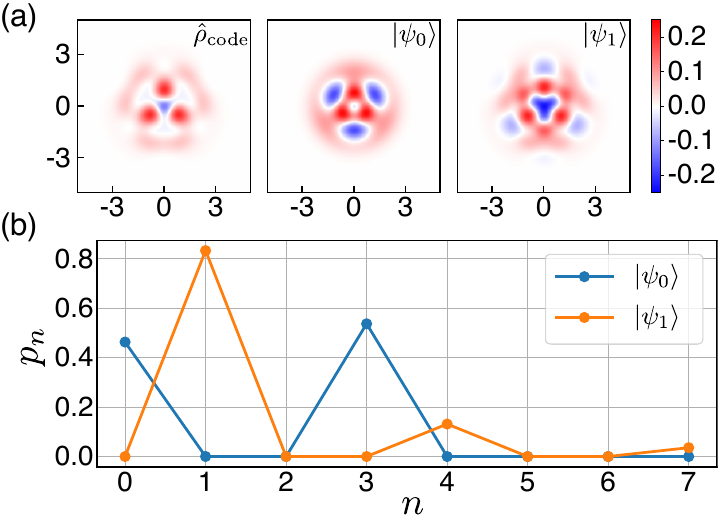}
    \caption{(a-b) Wigner functions and photon number distributions for another variant of the $\sqrt{3}$ code discovered with $d=2$ Hamiltonian.}
    \label{fig_SI_2}
\end{figure}

\begin{figure}
    \centering
    \includegraphics[width=0.48\textwidth]{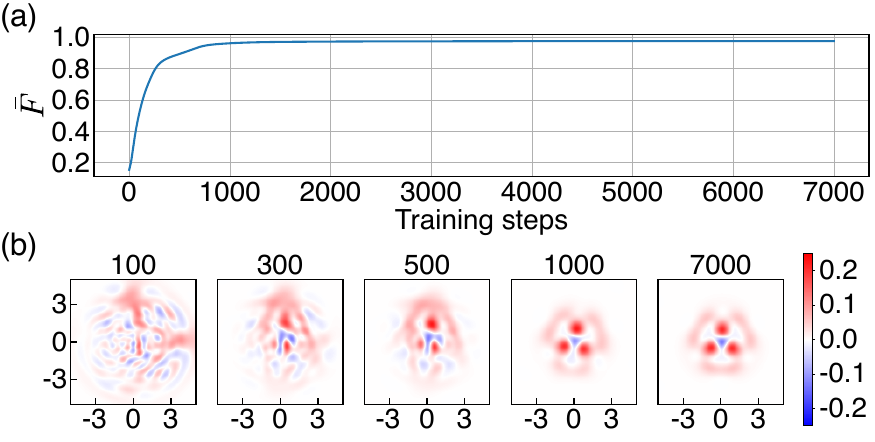}
    \caption{(a) Learning curve for results in Fig.~\ref{fig_SI_2}. (b) Wigner functions of $\hat{\rho}_{\text{code}}$ at different iterations during training, which shows a relatively good convergence after a few thousand iterations.}
    \label{fig_SI_3}
\end{figure}

\section{Additional comments on the optimization results}
\subsection{Exceed break-even with partial protection}
We further investigate the result in Fig.~\ref{fig2}(c)ii, which represents a class of optimization results that perform better than break-even fidelity but worse than full QEC codes.
Fig.~\ref{fig_SI_1}(a) shows the Wigner functions for the code subspace as well as both logical states, and Fig.\ref{fig_SI_1}(b) shows the photon number distribution for the logical states where $\ket{\psi_0} \in \{ \ket{1}, \ket{2}, \ket{3} \}$ only occupies low photon number states and $\ket{\psi_1} \in \{ \ket{5}, \ket{6}, \ket{7} \}$ only occupies high photon number states.

To understand how the logical subspace is preserved under the AQEC Hamiltonian, we plot the single state fidelity $F_{\theta\phi}$ over the Bloch sphere (Fig.~\ref{fig_SI_1}(c)) at $t=10~\mu$s. The logical state $\ket{\psi_0}$ is strongly stabilized by the AQEC Hamiltonian with a fidelity 0.985 and $\ket{\psi_1}$ is preserved with a lower fidelity 0.598.
Some of their superposition states ($\theta,\phi$ in-between the white dashed line) have fidelities below break-even, but the average fidelity over the whole Bloch sphere still exceeds break-even (Fig.~\ref{fig2}(e) red dashed line) due to the partial protection in the logical subspace.
In comparison, we also plot the single state fidelity for the $\sqrt{3}$ code (Fig.~\ref{fig2}(c)i) in Fig.~\ref{fig_SI_1}(d) which shows a relatively uniform protection for any logical states.

We could study a simplified example to demonstrate that a partially protected logical subspace exceeds break-even.
Stabilizing Fock states $\ket{0}$ and $\ket{2}$ under photon loss error can be implemented with a distance 1 Hamiltonian $\hat{H}=\ket{2,e} \bra{1,g}+\ket{1,g} \bra{2,e}$.
At long time, both logical states are stabilized with single state fidelities $F_{\theta=0}(t) \approx F_{\theta=\pi}(t) \approx 1$ but any coherent superposition state becomes a complete mixture of $\{\ket{0}\bra{0}, \ket{2}\bra{2}\}$.
This leads to an average fidelity of $\frac{2}{3}$ which is better than the break-even fidelity $\frac{1}{2}$.
Intuitively, stabilizing both $\ket{\psi_0}$ and $\ket{\psi_1}$ preserves strictly more information compared to collapsing the whole Bloch sphere to $\ket{\psi_0}=\ket{0}$.

\subsection{A different $\sqrt{3}$ code}
Besides the $\sqrt{3}$ code explained in the main text, \texttt{AutoQEC} also discovered another variant of $\sqrt{3}$ code (Fig.~\ref{fig_SI_2}(a)) protected by a distance 2 Hamiltonian.
The main difference is that $\ket{\psi_1} \in \{ \ket{1}, \ket{4}, \ket{7} \}$ instead of $\{ \ket{1}, \ket{4}, \ket{6} \}$ (Fig.~\ref{fig_SI_2}(b)). Following the same procedures as in Appendix~\ref{appendix:b}, this new code can also be analytically derived as ($F \approx 99.8\%$ compared to the numerical results in Fig.~\ref{fig_SI_2}(a))
\begin{equation}
    \begin{split}
        \ket{\psi_0} =& \sqrt{1-\frac{1}{\sqrt{3}}} \ket{0} + \frac{1}{\sqrt[4]{3}} \ket{3} \\
        \ket{\psi_1} =& \sqrt{\frac{4(7-\sqrt{3})}{3(7 + \sqrt{3})}} \ket{1} - \sqrt{\frac{(\sqrt{3}-1)(7-\sqrt{3})}{3(7 + \sqrt{3})}} \ket{4} \\
        & + \sqrt{\frac{3-\sqrt{3}}{3(7 + \sqrt{3})}} \ket{7} .
    \end{split}
\end{equation}

\section{Training details}
We use Adam optimizer~\cite{Kingma2015} for the gradient based learning with a learning rate about 0.001. Usually after a few hundred iterations, we can tell whether the training is stuck at a bad local minimum below break-even or not and make a decision on early stops. The training often achieves good convergence after a few thousand iterations and we could lower the learning rate to about 0.0003 for the final learning stage.

We choose a Fock state cutoff of 20 for the bosonic mode with a total Hilbert space dimension of 40. At the beginning of each \texttt{AutoQEC} run, the real and imaginary parts of the logical states $\ket{\psi_0}$ and $\ket{\psi_1}$ as length 40 complex vectors are randomly initialized.
During the optimization, in general $\ket{\psi_0}$ and $\ket{\psi_1}$ won't be perfectly orthogonal to each other after an Adam update step and therefore we choose to maintain their orthogonality by manually setting $\ket{\psi_1} \rightarrow \ket{\psi_1} - \frac{\inp{\psi_0}{\psi_1}}{\inp{\psi_0}{\psi_0}} \ket{\psi_0}$ after each iteration.

Figure~\ref{fig_SI_3} shows the learning curve for results in Fig.~\ref{fig_SI_2} discovered with $d=2$ Hamiltonian. Similar learning curves occur frequently through many runs of \texttt{AutoQEC}.
Regarding computational cost, each iteration takes about 12 seconds on 3 CPUs (Intel Xeon CPU E5-2609 v4 @ 1.70GHz) for training with distance 2 Hamiltonians.
\texttt{AutoQEC} runs on 3 CPUs because $\hat{\rho}_{00}(t),\hat{\rho}_{11}(t),\hat{\rho}_{10}(t)$ in the definition of $\bar{F}(t)$ can be evaluated in parallel with three independent master equation time evolutions.

\bibliography{LINQS_AutoQEC.bib}

\end{document}